\documentclass[aps,onecolumn,10pt]{revtex4}
\usepackage{epsfig}
\usepackage{graphicx}
\usepackage{amsmath}
\usepackage{amssymb}
\usepackage{hyperref}
\hypersetup{colorlinks=true,linkcolor=red,citecolor=green,urlcolor=blue}
\numberwithin{equation}{section}
\numberwithin{equation}{section}

\begin{document}
\allowdisplaybreaks
\setcounter{equation}{0}
\baselineskip=0.999cm
 
\title{Pauli-Villars  and the ultraviolet completion of Einstein gravity}
\author{Philip D. Mannheim}
\affiliation{Department of Physics, University of Connecticut, Storrs, CT 06269, USA \\
philip.mannheim@uconn.edu\\ }

\date{March 28 2023}

\begin{abstract}
Through use of the Pauli-Villars regulator procedure we construct a second- plus fourth-order-derivative theory of gravity that serves as an  ultraviolet completion of standard second-order-derivative quantum Einstein gravity that is ghost-free, unitary and power counting renormalizable. 

\end{abstract}

\maketitle
Essay written for the Gravity Research Foundation 2024 Awards for Essays on Gravitation.

\section{The Pauli-Villars regulator procedure}
\label{S1}

In order to regulate Feynman diagrams Pauli and Villars \cite{Pauli1949} suggested that one augment a propagating field (generically $\phi_1$) with a mirror field ($\phi_2$) that obeys all the same field equations  as $\phi_1$ but with a different mass and that is quantized with an indefinite (ghost state) metric. Taking the two fields together yields the free two-field action and quantization conditions 
\begin{align}
&I_1+I_2=\frac{1}{2}\displaystyle{\int}d^4x \left[\partial_{\mu}\phi_1\partial^{\mu}\phi_1-M_1^2\phi_1^2\right]
+\frac{1}{2}\displaystyle{\int}d^4x \left[\partial_{\mu}\phi_2\partial^{\mu}\phi_2-M_2^2\phi_2^2\right],
\nonumber\\
&[\phi_1(\bar{x},t),\dot{\phi}_1(\bar{x}^{\prime},t)]=i\delta^3(\bar{x}-\bar{x}^{\prime}),\qquad [\phi_2(\bar{x},t),\dot{\phi}_2(\bar{x}^{\prime},t)]=-i\delta^3(\bar{x}-\bar{x}^{\prime}),
\label{PV1.1}
\end{align}
and serves to replace a $1/(k^2-M_1^2+i\epsilon)$ propagator by 
\begin{align}
D(k)=\frac{1}{k^2-M_1^2+i\epsilon}-\frac{1}{k^2-M_2^2+i\epsilon}.
\label{PV1.2}
\end{align}
If the $\phi_1$ field propagator leads to a Feynman diagram that diverges as $\log[\Lambda^2/M_1^2]$, then the joint $\phi_1$, $\phi_2$ propagator $D(k)$ leads to a cut-off independent $\log[M_2^2/M_1^2]$. Prior to the Pauli-Villars paper  the concept of an indefinite metric had been discussed by Dirac \cite{Dirac1942} and Pauli \cite{Pauli1943} as a potential mechanism for controlling divergences in quantum field theory. While Pauli and Villars acknowledged in their paper that their use of an indefinite metric was only a mathematical device, they indicated that they did not want to rule out the possibility that it might be physical.  

To get further insight into this $D(k)$ propagator Pais and Uhlenbeck \cite{Pais1950} noted that this same $D(k)$  could be derived from a single field $\phi$ that obeyed a second- plus fourth-order theory action and equation of motion of the form
\begin{align}
I_S&=\frac{1}{2}\int d^4x\bigg{[}\partial_{\mu}\partial_{\nu}\phi\partial^{\mu}
\partial^{\nu}\phi-(M_1^2+M_2^2)\partial_{\mu}\phi\partial^{\mu}\phi
+M_1^2M_2^2\phi^2\bigg{]},\quad (\partial_t^2-\bar{\nabla}^2+M_1^2)(\partial_t^2-\bar{\nabla}^2+M_2^2)
\phi(x)=0.
\label{PV1.3}
\end{align}
For (\ref{PV1.3}) the associated propagator obeys
\begin{align}
&(\partial_t^2-\bar{\nabla}^2+M_1^2)(\partial_t^2-\bar{\nabla}^2+M_2^2)D(x)=-\delta^4(x),
\nonumber\\
&D(x)=-\int \frac{d^4k}{(2\pi)^4}\frac{e^{-ik\cdot x}}{(k^2-M_1^2+i\epsilon)(k^2-M_2^2+i\epsilon)}
=-\int \frac{d^4k}{(2\pi)^4}\frac{e^{-ik\cdot x}}{(M_1^2-M_2^2)}\left[\frac{1}{(k^2-M_1^2+i\epsilon)}-\frac{1}{(k^2-M_2^2+i\epsilon)}\right],
\label{PV1.4}
\end{align}
to thus have the same structure as given in (\ref{PV1.2}). For the $I_S$ action 
the energy-momentum tensor $T_{\mu\nu}$, the canonical momenta $\pi^{\mu}$ and $\pi^{\mu\lambda}$, and the equal-time commutators appropriate to the higher-derivative theory are given by \cite{Bender2008b} 
\begin{align}
T_{\mu\nu}&=\pi_{\mu}\phi_{,\nu}+\pi_{\mu}^{~\lambda}\phi_{,\nu,\lambda}-\eta_{\mu\nu}{\cal L},
\nonumber\\ 
 \pi^{\mu}&=\frac{\partial{\cal L}}{\partial \phi_{,\mu}}-\partial_{\lambda
}\left(\frac{\partial {\cal L}}{\partial\phi_{,\mu,\lambda}}\right)=-\partial_{\lambda}\partial^{\mu}\partial^{\lambda}\phi- (M_1^2+M_2^2)\partial^{\mu}\phi,\qquad  \pi^{\mu\lambda}=\frac{\partial {\cal L}}{\partial \phi_{,\mu,\lambda}}=\partial^{\mu}\partial^{\lambda}\phi,
\nonumber\\
T_{00}&=\tfrac{1}{2}\pi_{00}^2+\pi_{0}\dot{\phi}+\tfrac{1}{2}(M_1^2+M_2^2)\dot{
\phi}^2-\tfrac{1}{2}M_1^2M_2^2\phi^2
-\tfrac{1}{2}\pi_{ij}\pi^{ij}+\tfrac{1}{2}(M_1^2+M_2^2)\phi_{,i}\phi^{,i}
\nonumber\\
&=\frac{1}{2}\ddot{\phi}^2-\tfrac{1}{2}(M_1^2+M_2^2)\dot{
\phi}^2-\dddot{\phi}\dot{\phi}-[\partial_i\partial^i\dot{\phi}]\dot{\phi}
-\tfrac{1}{2}M_1^2M_2^2\phi^2
-\tfrac{1}{2}\partial_i\partial_j\phi\partial^i\partial^j\phi+\tfrac{1}{2}(M_1^2+M_2^2)\partial_i\phi\partial^i\phi,
\nonumber\\
&[\phi(\bar{0},t),\dot{\phi}(\bar{x},t)]=0, \qquad[\phi(\bar{0},t),\ddot{\phi}(\bar{x},t)]=0, \qquad [\phi(\bar{0},t),\dddot{\phi}(\bar{x},t])=-i\delta^3(x).
\label{PV1.5}
\end{align}
In terms of two sets of oscillator creation and annihilation operators the field  $\phi$ can be expressed as 
\begin{eqnarray}
\phi(\bar{x},t)=\int \frac{d^3k}{(2\pi)^{3/2}}\left[a_1(\bar{k})e^{-i\omega_1 t+i\bar{k}\cdot\bar{x}}+a^{\dagger}_1(\bar{k})e^{i\omega_1  t-i\bar{k}\cdot\bar{x}}+a_2(\bar{k})e^{-i\omega_2 t+i\bar{k}\cdot\bar{x}}+a^{\dagger}_2(\bar{k})e^{i\omega_2  t-i\bar{k}\cdot\bar{x}}\right],
\label{PV1.6}
\end{eqnarray}
where $\omega_1=+(\bar{k}^2+M_1^2)^{1/2}$, $\omega_2=+(\bar{k}^2+M_2^2)^{1/2}$. Given (\ref{PV1.6}) and the commutators and energy-momentum tensor given in (\ref{PV1.5}) we obtain a two-oscillator model of the form 
\begin{align}
H_S&=(M_1^2-M_2^2)\int d^3k\bigg{[}(\bar{k}^2+M_1^2)\left[a^{\dagger}_{1}(\bar{k})a_1(\bar{k})
+a_{1}(\bar{k})a^{\dagger}_1(\bar{k})\right]
-(\bar{k}^2+M_2^2)\left[a^{\dagger}_2(\bar{k})a_{2}(\bar{k})
+a_{2}(\bar{k})a^{\dagger}_2(\bar{k})\right]\bigg{]},
\nonumber\\
&[a_1(\bar{k}),a^{\dagger}_{1}(\bar{k}^{\prime})]=[2(M_1^2-M_2^2)(\bar{k}^2+
M_1^2)^{1/2}]^{-1}\delta^3(\bar{k}-\bar{k}^{\prime}),
\nonumber\\
&[a_2(\bar{k}),a^{\dagger}_{2}(\bar{k}^{\prime})]=-[2(M_1^2-M_2^2)(\bar{k}^2+
M_2^2)^{1/2}]^{-1}\delta^3(\bar{k}-\bar{k}^{\prime}),
\nonumber\\
&[a_1(\bar{k}),a_{2}(\bar{k}^{\prime})]=0,\quad[a_1(\bar{k}),a^{\dagger}_{2}(\bar{k}^{\prime})]=0,\quad[a^{\dagger}_1(\bar{k}),a_{2}(\bar{k}^{\prime})]=0,\quad
[a^{\dagger}_1(\bar{k}),a^{\dagger}_{2}(\bar{k}^{\prime})]=0.
\label{PV1.7}
\end{align}
We note that with $M_1^2-M_2^2>0$ for definitiveness, we see negative signs in both $H_S$ and the $[a_2(\bar{k}),a^{\dagger}_{2}(\bar{k}^{\prime})]$ commutator.  These negative signs signal two potential problems: states with negative norm (ghost states) and states with negative energy (the Ostrogradski instability that is characteristic of higher-derivative theories). However, these two problems cannot occur simultaneously as they occur in different Hilbert spaces. Specifically, one can define a Hilbert space built on a vacuum that $a_2(\bar{k})$ annihilates, or one can define a Hilbert space built on a vacuum that $a_2^{\dagger}(\bar{k})$ annihilates. In the first Hilbert space energies are positive but there are states of negative Dirac norm ($\langle \Omega|a_2(\bar{k})a_2^{\dagger}(\bar{k})|\Omega\rangle <0$), while in the second Hilbert space norms are positive but there are states of negative energy. The first Hilbert space corresponds to the standard Feynman $i\epsilon$ prescription given in (\ref{PV1.2}) and (\ref{PV1.4}) in which positive frequency modes propagate forward in time but the $M_2$ sector poles are enclosed in a way that leads to negative residues (closing in the lower half plane in the complex $k^0$ plane). The second Hilbert space corresponds to an $i\epsilon$ prescription of the form
\begin{align}
D^{\prime}(k)=\frac{1}{k^2-M_1^2+i\epsilon}-\frac{1}{k^2-M_2^2-i\epsilon}
\label{PV1.8}
\end{align}
in which negative energy states of energies up to minus infinity propagate forward in time, but the way that the $M_2$ sector  poles are enclosed leads to positive residues (closing in the upper half plane). Since it is not physically possible to have an energy spectrum that is unbounded from below, we shall work in the Hilbert space in which the energy spectrum is bounded from below. In this Hilbert space there is the issue of negative Dirac norms, but, as we shall describe below, this problem has been resolved  by Bender and Mannheim \cite{Bender2008a,Bender2008b} using the techniques of Bender's $PT$-symmetry program  \cite{Bender2007,Bender2019}. Interestingly the technique involves a continuation of the quantum fields into the complex plane that is the reverse of the one  that  Pauli used in order to construct a negative definite Hilbert space norm. Specifically, in \cite{Pauli1943} Pauli continued a positive definite norm into a negative definite norm, while in \cite{Bender2008a,Bender2008b} a negative definite norm is continued into a positive definite one. 

The need to continue into the complex plane is due to the fact that the Dirac norm $\langle \Omega\vert a_2(\bar{k})a_2^{\dagger}(\bar{k})\vert\Omega\rangle$ is not just negative, it turns out that it is infinite \cite{Bender2008a,Bender2008b,Mannheim2022,Mannheim2023a,Mannheim2023b,Mannheim2023c}, and thus cannot be the appropriate inner product for the quantum theory since while the residues of the some of the poles in (\ref{PV1.4}) may be negative, all of the residues, including the ones that are positive, are finite \cite{footnotePV1}. It is the infiniteness of the Dirac norm that actually saves the theory, since the very fact that the Dirac norm is infinite entails that the Hamiltonian is not self-adjoint, and thus not Hermitian. But all the $k^0$ plane poles in the propagator lie on the real $k^0$ axis, and thus despite the lack of Hermiticity the energy eigenvalues are nonetheless real. Now while Hermiticity implies reality of eigenvalues, there is no converse theorem that says that if a Hamiltonian is not Hermitian then it must have at least one complex eigenvalue. Hermiticity is only sufficient for reality but not necessary. A necessary condition has been identified in the literature, namely that the Hamiltonian possess an antilinear symmetry \cite{Bender2010,Mannheim2018}. And in such a case one should take the left eigenvector bra $\langle L\vert$ to be the conjugate  of a  right eigenvector ket $\vert R\rangle$ with respect to the antilinear symmetry operator rather than the Hermitian conjugate \cite{footnotePV2}. To resolve the infinity problem we need to continue the fields into a domain in the complex plane (known as a Stokes wedge) in which the Hamiltonian then is self-adjoint (we show below that such a domain does exist). In general \cite{Mannheim2018}  the antilinear symmetry should be $CPT$ ($C$ is charge conjugation, $P$ is parity, and $T$ is time reversal). But with the gravitational field being $C$ even the required inner product should be $\langle \Omega^{[PT]}\vert\Omega\rangle$ where  $\langle \Omega^{[PT]}\vert=PT\vert\Omega\rangle$  \cite{footnotePV3}, just as in the $PT$ program of Bender and collaborators \cite{Bender2007,Bender2019}. 

The essential point here is that one cannot determine the structure of a q-number Hilbert space by looking at a c-number quantity such as a propagator. Rather, first  one has construct the underlying  Hilbert space, and only then construct the propagator as a matrix element of a time ordered product of fields in the appropriate vacuum. With $i\langle\Omega\vert T(\phi(x)\phi(0))\vert \Omega\rangle$ obeying
\begin{align}
 (\partial_t^2-\bar{\nabla}^2+M_1^2)(\partial_t^2-\bar{\nabla}^2+M_2^2)
i\langle\Omega\vert T(\phi(x)\phi(0))\vert \Omega\rangle=-\langle \Omega\vert \Omega\rangle \delta^4(x),
\label{PV1.9}
\end{align}
the very fact that $\langle \Omega\vert \Omega\rangle$ is infinite means that the propagator cannot be represented by 
$i\langle\Omega\vert T(\phi(x)\phi(0))\vert \Omega\rangle$. Rather, it has to be of the form $-i\langle \Omega^{[PT]}\vert T(\bar{\phi}(x)\bar{\phi}(0))\vert \Omega\rangle$, as based on the continuation  $\bar{\phi}=-i\phi$ of $\phi$ into the complex plane.  Thus the fact that the second- plus fourth-order theory is thought to have unitarity violating states of negative norm is due to a misidentification of what quantum field theory matrix element the propagator is to represent. It is in this way that the Pauli-Villars regulator can be physical, as can then also be the original program of Dirac and Pauli. 

Now as well as regulate logarithmically divergent Feynman diagrams the Pauli-Villars regulator can also regulate quadratically divergent ones as well and reduce them to renormalizable logarithmic divergences. Such quadratic divergences occur when the standard second-order Einstein gravity theory is quantized, and so we now show how we can now use a Pauli-Villars-based gravity theory to control them in a way that is unitary and ghost free.

\section{Quantum Einstein gravity}

While the Pauli-Villars propagator reduces a $1/k^2$ behavior at asymptotic $k^2$  to the more convergent $1/k^4$, and while radiative corrections to Einstein gravity do generate (Planck-scale) higher-derivative counter terms, they leave the asymptotic behavior as $1/k^2$, to thus render the theory nonrenormalizable. Thus if we want the Pauli-Villars propagator to be a physical component of a quantum gravity theory that is to control the asymptotic behavior we must introduce higher-derivative terms right at the beginning in the fundamental Lagrangian. To this end we augment the Einstein Ricci scalar action with a term that is quadratic in the Ricci scalar. This gives a much studied \cite{footnotePV4} quantum gravity action of the generic form
\begin{eqnarray}
I_{\rm GRAV}=\int d^4x(-g)^{1/2}\left[6M^2R^{\alpha}_{~\alpha}+(R^{\alpha}_{~\alpha})^2\right].
\label{PV2.1}
\end{eqnarray}
%

On adding on a matter source with energy-momentum tensor $T_{\mu\nu}$, variation  of this action with respect to the metric generates a gravitational equation of motion of the form
\begin{eqnarray}
-6M^2G^{\mu\nu}+V^{\mu\nu}=-\frac{1}{2}T^{\mu\nu}.
\label{PV2.2}
\end{eqnarray}
Here $G_{\mu\nu}$ is the Einstein tensor and $V_{\mu\nu}$ may  be found in \cite{DeWitt1964} and also \cite{Mannheim2006,Mannheim2017}, with these  terms being of the form 
\begin{align}
G^{\mu\nu}&=R^{\mu\nu}-\frac{1}{2}g^{\mu\nu}g^{\alpha\beta}R_{\alpha\beta},\quad
V^{\mu \nu}=
2g^{\mu\nu}\nabla_{\beta}\nabla^{\beta}R^{\alpha}_{~\alpha}                                             
-2\nabla^{\nu}\nabla^{\mu}R^{\alpha}_{~\alpha}                          
-2 R^{\alpha}_{\phantom{\alpha}\alpha}R^{\mu\nu}                              
+\frac{1}{2}g^{\mu\nu}(R^{\alpha}_{~\alpha})^2.
\label{PV2.3}
\end{align}                                 
If we now linearize about  flat spacetime with background metric $\eta_{\mu\nu}$ and fluctuation  metric $g_{\mu\nu}=\eta_{\mu\nu}+h_{\mu\nu}$, to first perturbative order we obtain 
\begin{align}
\delta G_{\mu\nu}&=\frac{1}{2}\left(\partial_{\alpha}\partial^{\alpha}h_{\mu\nu}-\partial_{\mu}\partial^{\alpha}h_{\alpha\nu}-\partial_{\nu}\partial^{\alpha}h_{\alpha\mu}+\partial_{\mu}\partial_{\nu}h\right)-\frac{1}{2}\eta_{\mu\nu}\left(\partial_{\alpha}\partial^{\alpha}h-\partial^{\alpha}\partial^{\beta}h_{\alpha\beta}\right),
\nonumber\\
\delta V_{\mu\nu}&=[2\eta_{\mu\nu}\partial_{\alpha}\partial^{\alpha} -2\partial_{\mu}\partial_{\nu}]
[\partial_{\beta}\partial^{\beta}h-\partial_{\lambda}\partial_{\kappa}h^{\lambda\kappa}],
\label{PV2.4}
\end{align}                                 
where $h=\eta^{\mu\nu}h_{\mu\nu}$. Since all the $h_{\mu\nu}$ fluctuation components diverge at the same rate, we only need to look at one of them in order to see the general pattern, with the trace $h$ being the most convenient. Thus taking the trace of the fluctuation equation around a flat background we obtain 
\begin{eqnarray}
[M^2+\partial_{\beta}\partial^{\beta}]\left(\partial_{\lambda}\partial^{\lambda}h-\partial_{\kappa}\partial_{\lambda}h^{\kappa\lambda}\right)=-\frac{1}{12}\eta^{\mu\nu}\delta T_{\mu\nu}.
\label{PV2.5}
\end{eqnarray}
In the convenient transverse gauge where $\partial_{\mu}h^{\mu\nu}=0$, the propagator for $h$ is given by
\begin{eqnarray}
D(h,k^2)=-\frac{1}{(k^2+i\epsilon)(k^2-M^2+i\epsilon)}=\frac{1}{M^2}\left(\frac{1}{k^2+i\epsilon}-\frac{1}{k^2-M^2+i\epsilon}\right).
\label{PV2.6}
\end{eqnarray}
As we see, in this case the $1/k^2$ graviton propagator for $h$ that would be associated with the Einstein tensor $\delta G_{\mu\nu}$ alone  is replaced by a far more convergent $D(h,k^2)=[1/k^2-1/(k^2-M^2)]/M^2$ propagator. And now the leading behavior at large momenta is $-1/k^4$. If we identify this propagator with $i\langle \Omega \vert T(h(x)h(0))\vert \Omega$, and assume that $\langle \Omega\vert\Omega \rangle$ is finite the Feynman rules that would then ensue render the theory renormalizable \cite{Stelle1977}. But since, as we show below,   $\langle \Omega\vert\Omega \rangle$ is not finite the proof of renormalizability has a flaw in it. Fortunately, the flaw is not fatal, and we rectify it below, with $\langle \Omega^{[PT]}\vert\Omega \rangle$ being finite and ghost free \cite{footnotePV3} and the well-behaved $-i\langle \Omega^{[PT]}\vert T(\bar{h}(x)\bar{h}(0))\vert\Omega\rangle$ with $\bar{h}=-ih$ obeying the same Feynman rules, so that renormalizability is obtained.

We recognize $D(h,k^2)$ as being of the same form as the scalar field propagator that was given in (\ref{PV1.4}), with $\phi$ being replaced by $h$ and with $M_1^2=M^2$, $M_2^2=0$. We can thus give $h$ an equivalent effective action of the form
\begin{eqnarray}
I_h&=&\frac{1}{2}\int d^4x\bigg{[}\partial_{\mu}\partial_{\nu}h\partial^{\mu}
\partial^{\nu}h-M^2\partial_{\mu}h\partial^{\mu}h\bigg{]}.
\label{PV2.7}
\end{eqnarray}
$I_h$  thus shares the same vacuum normalization and negative norm challenges as the scalar field action given in (\ref{PV1.3}).

To see whether this theory can be considered to be an ultraviolet completion of Einstein gravity, we need to be able to find a consistent Hilbert space formulation  that will realize this propagator, and this we now do.

\section{The inappropriate Hilbert space}

Since $I_h$ only involves one degree of freedom we can treat it as a scalar field $\phi$ and use the action given in (\ref{PV1.3}). To be general we shall keep both $M_1^2$ and $M_2^2$  arbitrary. To construct the appropriate Hilbert space we need to utilize the techniques presented in \cite{Bender2008a,Bender2008b,Mannheim2022,Mannheim2023a,Mannheim2023b,Mannheim2023c}. To this end we introduce operators
\begin{align}
z(\bar{k},t)&=a_1(\bar{k})e^{-i\omega_1(\bar{k})t}+a_1^{\dagger}(\bar{k})e^{i\omega_1(\bar{k})t}+a_2(\bar{k})e^{-i\omega_2(\bar{k})t}+a_2^{\dagger}(\bar{k})^{i\omega_2(\bar{k})t},
\nonumber\\
p_z(\bar{k},t)&=i\omega_1(\bar{k})\omega_2^2(\bar{k})
[a_1(\bar{k})e^{-i\omega_1(\bar{k})t}-a_1^{\dagger}(\bar{k})e^{i\omega_1(\bar{k})t}]+i\omega_1^2(\bar{k})\omega_2(\bar{k})[a_2(\bar{k})e^{-i\omega_2(\bar{k})t}-a_2^{\dagger}(\bar{k})e^{i\omega_2(\bar{k})t}],
\nonumber\\
x(\bar{k},t)&=-i\omega_1(\bar{k})[a_1(\bar{k})e^{-i\omega_1(\bar{k})t}-a_1^{\dagger}(\bar{k})e^{i\omega_1(\bar{k})t}]-i\omega_2(\bar{k})[a_2(\bar{k})e^{-i\omega_2(\bar{k})t}-a_2^{\dagger}(\bar{k})^{i\omega_2(\bar{k})t}],
\nonumber\\
p_x(\bar{k},t)&=-\omega_1^2(\bar{k})[a_1(\bar{k})e^{-i\omega_1(\bar{k})t}+a_1^{\dagger}(\bar{k})e^{i\omega_1(\bar{k})t}]-\omega_2^2(\bar{k})[a_2(\bar{k})e^{-i\omega_2(\bar{k})t}+a_2^{\dagger}(\bar{k})^{i\omega_2(\bar{k})t}].
\label{PV3.1}
\end{align}
Given the commutation relations that appear  in (\ref{PV1.7}) it follows that 
\begin{eqnarray}
&&[z(\bar{k},t),p_z(\bar{k}^{\prime},t)]=i\delta^3(\bar{k}-\bar{k}^{\prime}),\qquad [x(\bar{k},t),p_x(\bar{k}^{\prime},t)]=i\delta^3(\bar{k}-\bar{k}^{\prime}),
\nonumber\\
&&[z(\bar{k},t),x(\bar{k}^{\prime},t)]=0,\quad[z(\bar{k},t),p_x(\bar{k}^{\prime},t)]=0,\quad[p_z(\bar{k},t),x(\bar{k}^{\prime},t)]=0,\quad
[p_z(\bar{k},t),p_x(\bar{k}^{\prime},t)]=0,
\label{PV3.2}
\end{eqnarray}
Thus for each $\bar{k}$  we rewrite the commutation algebra in the standard position and momentum algebra form that is familiar in quantum mechanics. Similarly,  we can rewrite the Hamiltonian given in (\ref{PV1.7}) in the equivalent form
\begin{align}
H_S=\int d^3k\bigg{[}\frac{p_x^2(\bar{k},t)}{2}+p_z(\bar{k},t)x(\bar{k},t)+\frac{1}{2}\left[\omega_1^2(\bar{k})+\omega_2^2(\bar{k}) \right]x^2(\bar{k},t)-\frac{1}{2}\omega_1^2(\bar{k})\omega_2^2(\bar{k})z^2(\bar{k},t)\bigg{]}=\int d^3kH_{\rm PU}(\bar{k}).
\label{PV3.3}
\end{align}
The quantum field theory $H_S$ is diagonal in the $\bar{k}$ basis, and  for each $\bar{k}$ we recognize the quantum-mechanical  Pais-Uhlenbeck (PU) Hamiltonian $H_{\rm PU}(\bar{k})$  that was derived in \cite{Mannheim2000,Mannheim2005} and discussed in detail in \cite{Bender2008a,Bender2008b}.

We have chosen the particular basis for the commutation algebra that is given in (\ref{PV3.2}) because we can represent it in a differential form 
\begin{eqnarray}
&&\left[z(\bar{k},t), -i\frac{\partial}{\partial z(\bar{k}^{\prime},t)}\right]=\delta^3(\bar{k}-\bar{k}^{\prime}),\qquad \left[x(\bar{k},t),-i\frac{\partial}{\partial x(\bar{k}^{\prime},t)}\right]=\delta^3(\bar{k}-\bar{k}^{\prime})
\label{PV3.4}
\end{eqnarray}
that will enable us to construct wave functions and explore their asymptotic behavior.
With the vacuum obeying $a_1(\bar{k})\vert \Omega\rangle=0$, $a_2(\bar{k})\vert \Omega\rangle=0$ for each $\bar{k}$, from (\ref{PV3.4})  we obtain 
\begin{align}
&\langle z(\bar{k}),x(\bar{k})\vert a_1(\bar{k})\vert \Omega\rangle=\frac{1}{2(M_1^2-M_2^2)}\left[-\omega_2^2(\bar{k})z(\bar{k})+i\frac{\partial}{\partial x(\bar{k})}+i\omega_1(\bar{k})x(\bar{k})+\frac{1}{\omega_1(\bar{k})}\frac{\partial}{\partial z(\bar{k})}\right]\langle z(\bar{k}),x(\bar{k})\vert\Omega\rangle=0,
\nonumber\\
&\langle z(\bar{k}),x(\bar{k})\vert a_2(\bar{k})\vert \Omega\rangle=\frac{1}{2(M_1^2-M_2^2)}\left[\omega_1^2(\bar{k})z(\bar{k})-i\frac{\partial}{\partial x(\bar{k})}-i\omega_2(\bar{k})x(\bar{k})-\frac{1}{\omega_2(\bar{k})}\frac{\partial}{\partial z(\bar{k})}\right]\langle z(\bar{k}),x(\bar{k})\vert\Omega\rangle=0,
\label{PV3.5}
\end{align}
for each $\bar{k}$. From (\ref{PV3.5}) we identify the ground state wave function $\psi_0(z(\bar{k}),x(\bar{k}))=\langle z(\bar{k}),x(\bar{k})\vert\Omega\rangle$ to be of the form
\begin{align}
\psi_0(z(\bar{k}),x(\bar{k}))=\exp[\tfrac{1}{2}[\omega_1(\bar{k})+\omega_2(\bar{k})]\omega_1(\bar{k})\omega_2(\bar{k})z^2(\bar{k})+i\omega_1(\bar{k})\omega_2(\bar{k})z(\bar{k})x(\bar{k})-\tfrac{1}{2}[\omega_1(\bar{k})+\omega_2(\bar{k})]x^2(\bar{k})].
\label{PV3.6}
\end{align}
Consequently,  the Dirac norm of the vacuum of the full $H_S$ is given by
\begin{align}
\langle \Omega\vert \Omega\rangle&=\Pi_{\bar{k}}\int_{-\infty}^{\infty} dz(\bar{k})\int_{-\infty}^{\infty}dx(\bar{k})\langle \Omega\vert  z(\bar{k}),x(\bar{k})\rangle\langle z(\bar{k}),x(\bar{k})\vert\Omega\rangle
\nonumber\\
&=\Pi_{\bar{k}}\int_{-\infty}^{\infty} dz(\bar{k}) \int_{-\infty}^{\infty}dx(\bar{k}) \psi_0^*(z(\bar{k}),x(\bar{k}))\psi_0(z(\bar{k}),x(\bar{k})).
\label{PV3.7}
\end{align}
With each $\psi_0(z(\bar{k}),x(\bar{k}))$ diverging at large $z(\bar{k})$, we thus establish that the Dirac norm of the field theory vacuum is infinite. Since the $\psi_0(z(\bar{k}),x(\bar{k}))$ would not diverge if we continued each $z(\bar{k})$ into the complex plane according to $z(\bar{k})\rightarrow -iz(\bar{k})=y(\bar{k})$ (and accordingly  $p_z(\bar{k})\rightarrow ip_z(\bar{k})=q(\bar{k})$ for the canonical conjugate) we would then secure finiteness. As we now, this will also provide us with an inner product that is  positive.

\section{The appropriate Hilbert space}

Since $\psi_0(z(\bar{k}),x(\bar{k}))$ converges at large $x(\bar{k})$ we have no need to continue  $x(\bar{k})$ into the complex plane, i.e., we only need to continue one of the two oscillators (viz. $z(\bar{k})$) and not the other. To implement the continuation we make the similarity transformation  $\phi(\bar{x},t)\rightarrow  S\phi S^{-1}=-i\phi(\bar{x},t)=\bar{\phi}(\bar{x},t)$, where  $S=\exp[\pi \int d^3x\pi_0(\bar{x},t)\phi(\bar{x},t)/2]$, and expand $\bar{\phi}(\bar{x},t)$ in a complete basis of plane waves as \cite{footnotePV5}
\begin{align}
\bar{\phi}(x)&=\int \frac{d^3k}{(2\pi)^{3/2}}\left [-ia_1(\bar{k})e^{-i\omega_1(\bar{k})t+i\bar{k}\cdot \bar{x}}+a_2(\bar{k})e^{-i\omega_2(\bar{k}) t+i\bar{k}\cdot \bar{x}}-i\hat{a}_1(\bar{k})e^{i\omega_1(\bar{k}) t-i\bar{k}\cdot \bar{x}}+\hat{a}_2(\bar{k})e^{i\omega_2(\bar{k}) t-i\bar{k}\cdot \bar{x}}\right].
\label{PV4.1}
\end{align}
Given (\ref{PV4.1}),  the transformed Hamiltonian $\bar{H}_S$ and transformed commutators now take the form \cite{Bender2008b}
\begin{align}\bar{H}_S&=(M_1^2-M_2^2)\int d^3k\bigg{[}(\bar{k}^2+M_1^2)
\left[\hat{a}_{1}(\bar{k})a_1(\bar{k})+a_{1}(\bar{k})\hat{a}_1(\bar{k})\right]
+(\bar{k}^2+M_2^2)\left[\hat{a}_{2}(\bar{k})a_2(\bar{k})+a_{2}(\bar{k})\hat{a}_2(\bar{k})\right]\bigg{]},
\nonumber\\
& [\dot{\bar{\phi}}(\bar{x},t),\bar{\phi}(0)]=0,\qquad [\ddot{\bar{\phi}}(\bar{x},t),\bar{\phi}(0)]=0,\qquad [\dddot{\bar{\phi}}(\bar{x},t),\bar{\phi}(0)]=i\delta^3(x),
\nonumber\\
&[a_1(\bar{k}),\hat{a}_{1}(\bar{k}^{\prime})]=[2(M_1^2-M_2^2)(\bar{k}^2+
M_1^2)^{1/2}]^{-1}\delta^3(\bar{k}-\bar{k}^{\prime}),
\nonumber\\
&[a_2(\bar{k}),\hat{a}_{2}(\bar{k}^{\prime})]=[2(M_1^2-M_2^2)(\bar{k}^2+
M_2^2)^{1/2}]^{-1}\delta^3(\bar{k}-\bar{k}^{\prime}),
\nonumber\\
&[a_1(\bar{k}),a_{2}(\bar{k}^{\prime})]=0,
\quad [a_1(\bar{k}),\hat{a}_{2}(\bar{k}^{\prime})]=0,
\quad [\hat{a}_{1}(\bar{k}),a_{2}(\bar{k}^{\prime})]=0,
\quad [\hat{a}_{1}(\bar{k}),\hat{a}_{2}(\bar{k}^{\prime})]=0,
\label{PV4.2}
\end{align}
The algebra of the creation and annihilation operators as given in (\ref{PV4.2}) provides a faithful representation of the field commutation relations. With all relative signs in (\ref{PV4.2}) being positive (we take $M_1^2>M_2^2$), there are no states of negative norm or of negative energy, and the theory is now fully viable. As such, this discussion extends to field theory the previously published discussion of the quantum-mechanical PU two-oscillator model that was given in \cite{Bender2008a,Bender2008b}.

To complete the field theory discussion we replace (\ref{PV3.1}), (\ref{PV3.2}) and (\ref{PV3.3}) by 
\begin{align}
y(\bar{k},t)&=-ia_1(\bar{k})e^{-i\omega_1(\bar{k})t}+a_2(\bar{k})e^{-i\omega_2(\bar{k})t}-i\hat{a}_1(\bar{k})e^{i\omega_1(\bar{k})t}+\hat{a}_2(\bar{k})
e^{i\omega_2(\bar{k})t},
\nonumber\\
x(\bar{k},t)&=-i\omega_1(\bar{k})a_1(\bar{k})e^{-i\omega_1(\bar{k})t}+\omega_2(\bar{k})a_2(\bar{k})e^{-i\omega_2(\bar{k})t}+i\omega_1(\bar{k})\hat{a}_1(\bar{k})
e^{i\omega_1(\bar{k})t}-\omega_2(\bar{k})\hat{a}_2(\bar{k})e^{i\omega_2(\bar{k})t},
\nonumber\\
p_x(\bar{k},t)&=-\omega_1^2(\bar{k})a_1(\bar{k})e^{-i\omega_1(\bar{k})t}-i\omega_2^2(\bar{k})a_2(\bar{k})e^{-i\omega_2(\bar{k})t}-\omega_1
^2(\bar{k})\hat{a}_1(\bar{k})e^{i\omega_1(\bar{k})t}-i\omega_2^2(\bar{k})\hat{a}_2(\bar{k})e^{i\omega_2(\bar{k})t},
\nonumber\\
q(\bar{k},t)&=\omega_1(\bar{k})\omega_2(\bar{k})[-\omega_2(\bar{k})a_1(\bar{k})e^{-i\omega_1(\bar{k})t}-i\omega_1(\bar{k})a_2(\bar{k})e^{-i\omega_2(\bar{k})t}+\omega_2(\bar{k})\hat{a}_1(\bar{k})e^{i\omega_1(\bar{k})t}+i\omega_1(\bar{k})\hat{a}_2(\bar{k})e^{i\omega_2(\bar{k})t}],
\nonumber\\
&[y(\bar{k},t),q(\bar{k}^{\prime},t)]=i\delta^3(\bar{k}-\bar{k}^{\prime}),\qquad [x(\bar{k},t),p_x(\bar{k}^{\prime},t)]=i\delta^3(\bar{k}-\bar{k}^{\prime}),
\nonumber\\
\bar{H}_S&=\int d^3k\bigg{[}\frac{p_x^2(\bar{k},t)}{2}-iq(\bar{k},t)x(\bar{k},t)+\frac{1}{2}\left[\omega_1^2(\bar{k})+\omega_2^2(\bar{k}) \right]x^2(\bar{k},t)+\frac{1}{2}\omega_1^2(\bar{k})\omega_2^2(\bar{k})y^2(\bar{k},t)\bigg{]}=\int d^3k\bar{H}_{\rm PU}(\bar{k}).
\label{PV4.3}
\end{align}
The factor of $i$ present in $\bar{H}_S$ shows that it is not is not Hermitian. Nonetheless, it still has all eigenvalues real since it is $PT$ symmetric \cite{Bender2008a,Mannheim2023b} and since its real eigenvalues anyway cannot change under a similarity transformation.

Now when a Hamiltonian is not Hermitian the action of it to the right and the action of it to the left are not related by Hermitian conjugation. Thus in general one must distinguish between right- and left-eigenstates (the left-eigenstates are the same as the $PT$ conjugated right-eigenstates), both for the vacuum and the states that can be excited out of it, and one must use the left-right inner product. This inner product obeys $\langle L(t)\vert R(t)\rangle=\langle L(0)\vert e^{iHt}e^{-iHt}\vert R(0)\rangle=\langle L(0)\vert R(0)\rangle$, to thus nicely be time independent in the non-Hermitian case. In the left-right basis we represent the equal-time $[y(\bar{k}),q(\bar{k^{\prime}})]=i$ and  $[x(\bar{k}),p_x(\bar{k}^{\prime})]=i$ commutators by  $q(\bar{k}^{\prime})=-i\overrightarrow{\partial_y(\bar{k}^{\prime})}$, $p_x(\bar{k}^{\prime})=-i\overrightarrow{\partial_x(\bar{k}^{\prime})}$ when acting to the right, and by $q(\bar{k})=i\overleftarrow{\partial_y(\bar{k}^{\prime})}$, $p_x(\bar{k}^{\prime})=i\overleftarrow{\partial_x(\bar{k}^{\prime})}$ when acting to the left. This  leads to right and left ground state wave functions of the form \cite{Bender2008b}
\begin{align}
\psi_0^R(y(\bar{k}),x(\bar{k}))&=\exp[-\tfrac{1}{2}(\omega_1(\bar{k})+\omega_2(\bar{k}))\omega_1(\bar{k})\omega_2(\bar{k})y^2(\bar{k})-\omega_1(\bar{k})\omega_2(\bar{k})y(\bar{k})x(\bar{k})-\tfrac{1}{2}(\omega_1(\bar{k})+\omega_2(\bar{k}))x^2(\bar{k})],
\nonumber\\
\psi_0^L(y(\bar{k}),x(\bar{k}))&=\exp[-\tfrac{1}{2}(\omega_1(\bar{k})+\omega_2(\bar{k}))\omega_1(\bar{k})\omega_2(\bar{k})y^2(\bar{k})+\omega_1(\bar{k})\omega_2(\bar{k})y(\bar{k})x(\bar{k})-\tfrac{1}{2}(\omega_1(\bar{k})+\omega_2(\bar{k}))x^2(\bar{k})].
\label{PV4.4}
\end{align}
Given these wave functions  the vacuum normalization for each $\bar{k}$ is given by \cite{Bender2008b}
 \begin{align}
 \langle \Omega^{L}(\bar{k})\vert \Omega^R(\bar{k})\rangle&=\int_{-\infty}^{\infty} dy(\bar{k})\int_{-\infty}^{\infty}dx(\bar{k})\langle \Omega^{L}\vert y(\bar{k}),x(\bar{k})\rangle\langle y(\bar{k}),x(\bar{k})\vert\Omega^R\rangle
 \nonumber\\
 &=\int_{-\infty}^{\infty} dy(\bar{k})\int_{-\infty}^{\infty}dx(\bar{k})\psi_0^L(y(\bar{k}),x(\bar{k}))\psi_0^R(y(\bar{k}),x(\bar{k}))
 =\frac{\pi}{(\omega_1(\bar{k})\omega_2(\bar{k}))^{1/2}(\omega_1(\bar{k})+\omega_2(\bar{k}))}.
 \label{PV4.5}
 \end{align}
Thus, just as we want, the left-right inner product normalization is finite. On normalizing each $ \langle \Omega^{L}(\bar{k})\vert \Omega^R(\bar{k})\rangle$ to one we find that 
\begin{align}
\langle \Omega^{L}\vert \Omega^R\rangle&=\Pi_{\bar{k}} \langle \Omega^{L}(\bar{k})\vert \Omega^R(\bar{k})\rangle=
\Pi_{\bar{k}}1=1.
\label{PV4.6}
\end{align}
We thus confirm that the left-right vacuum normalization is both finite and positive. We thus establish the consistency and physical viability of the similarity-transformed higher-derivative scalar field theory. And we note that even though all the norms are  positive, the insertion into $-i\langle\Omega^{L}\vert T[\bar{\phi}(x)\bar{\phi}(0)]\vert \Omega^R \rangle$ (corresponding to  $+i\langle\Omega^{L}\vert T[\phi(x)\phi(0)]\vert \Omega^R \rangle$) of $\bar{\phi}(x)$ with its explicit $i$ factors as given in (\ref{PV4.1})  generates \cite{Bender2008b} the relative minus sign in $-[1/(k^2-M_1^2)-1/(k^2-M_2^2)]/(M_1^2-M_2^2)$ \cite{footnotePV6}. Thus with one similarity transform into an appropriate Stokes wedge we solve both the vacuum normalization problem and the negative-norm problem. 

In the gravity case  with $\bar{h}=-ih$ the propagator is given by $-i\langle \Omega^{L}\vert T[\bar{h}(x)\bar{h}(0)]\vert \Omega^R \rangle$ (corresponding to $+i\langle \Omega^{L}\vert T[h(x)h(0)]\vert \Omega^R \rangle$). And with the propagator still being given by (\ref{PV2.6}) as it satisfies $(\partial_t^2-\bar{\nabla}^2+M^2)(\partial_t^2-\bar{\nabla}^2)[-i\langle \Omega^{L}\vert T[\bar{h}(x)\bar{h}(0)]\vert \Omega^R \rangle]=-\delta^4(x)$, all the steps needed to prove renormalizability are now valid.  Consequently, the theory can now be offered as a fully consistent, unitary and renormalizable theory of quantum gravity, one whose low energy ($k^2 \ll M^2$) limit is based on the standard $1/k^2$ propagator of quantum Einstein gravity \cite{footnotePV7}, \cite{footnotePV8}.


\begin{thebibliography}{99}
\bibitem{footnotePV1} It may be thought that if $M_2 \gg M_1$ one could avoid negative residue concerns at energies way below $M_2$,  to thereby permit the $M_1$ sector to be considered as an effective low energy theory. However this cannot be the case since even in the $M_1$ sector the Dirac norm is infinite, as would then be Green's functions such as $i\langle \Omega\vert T(\phi(x)\phi(0))\vert \Omega \rangle$  that would be based on it, being infinite even in the  low momentum limit.
\bibitem{footnotePV2} The freedom here is due to the fact that the in constructing eigenstates of a Hamiltonian $H$ according to $H\vert \psi\rangle=E\vert \psi \rangle$ only the ket is specified, not the bra.
\bibitem{footnotePV5} We use the hatted notation to indicate that the hatted quantities are not the Hermitian conjugates of the corresponding unhatted ones. Rather,  they are the $PT$ transformed ones \cite{Mannheim2023b}.
\bibitem{footnotePV6} The reason why the relative minus sign is generated is because the insertion of the generalized  closure relation $\sum \vert R\rangle\langle L \vert= I$ into the propagator does not give a modulus squared.
\bibitem{footnotePV7} Even though we have continued the theory into the complex plane, as noted in \cite{Mannheim2023b}, the ensuing classical limit is still real.
\end{thebibliography}
\end{document}